\documentclass{pasj01}

\RequirePackage{mathpazo}
\RequirePackage[T1]{fontenc}

\newcounter{author}
\def\authorcount#1#2{\refstepcounter{author}\label{#1}
                     \altaffiltext{\ref{#1}}{#2}}

\begin{document}
\SetRunningHead{T. Kato et al.}{IW And-Type State in IM Eridani}

\Received{201X/XX/XX}
\Accepted{201X/XX/XX}

\title{IW And-Type State in IM Eridani}

\author{Taichi~\textsc{Kato},\altaffilmark{\ref{affil:Kyoto}*}
        Yasuyuki~\textsc{Wakamatsu},\altaffilmark{\ref{affil:Kyoto}}
        Naoto~\textsc{Kojiguchi},\altaffilmark{\ref{affil:Kyoto}}
        Mariko~\textsc{Kimura},\altaffilmark{\ref{affil:Kyoto}}
        Ryuhei~\textsc{Ohnishi},\altaffilmark{\ref{affil:Kyoto}}
        Keisuke~\textsc{Isogai},\altaffilmark{\ref{affil:Kyoto}}$^,$\altaffilmark{\ref{affil:KyotoOkayama}}
        Keito~\textsc{Niijima},\altaffilmark{\ref{affil:Kyoto}}
        Tomohiro~\textsc{Yoshitake},\altaffilmark{\ref{affil:Kyoto}}
        Yuki~\textsc{Sugiura},\altaffilmark{\ref{affil:OKU}}
        Sho~\textsc{Sumiya},\altaffilmark{\ref{affil:OKU}}
        Daiki~\textsc{Ito},\altaffilmark{\ref{affil:OKU}}
        Kengo~\textsc{Nikai},\altaffilmark{\ref{affil:OKU}}
        Hanami~\textsc{Matsumoto},\altaffilmark{\ref{affil:OKU}}
        Katsura~\textsc{Matsumoto},\altaffilmark{\ref{affil:OKU}}
        Tonny~\textsc{Vanmunster},\altaffilmark{\ref{affil:Vanmunster}}
        Franz-Josef~\textsc{Hambsch},\altaffilmark{\ref{affil:GEOS}}$^,$\altaffilmark{\ref{affil:BAV}}$^,$\altaffilmark{\ref{affil:Hambsch}}
        Hiroshi~\textsc{Itoh},\altaffilmark{\ref{affil:Ioh}}
        Julia~V.~\textsc{Babina},\altaffilmark{\ref{affil:CrAO}}
        Oksana~I.~\textsc{Antonyuk},\altaffilmark{\ref{affil:CrAO}}
        Alex~V.~\textsc{Baklanov},\altaffilmark{\ref{affil:CrAO}}
        Elena~P.~\textsc{Pavlenko},\altaffilmark{\ref{affil:CrAO}}
        Berto~\textsc{Monard},\altaffilmark{\ref{affil:Monard}}$^,$\altaffilmark{\ref{affil:Monard2}}
        Shawn~\textsc{Dvorak},\altaffilmark{\ref{affil:Dvorak}}
}

\authorcount{affil:Kyoto}{
     Department of Astronomy, Kyoto University, Kyoto 606-8502, Japan}
\email{$^*$tkato@kusastro.kyoto-u.ac.jp}

\authorcount{affil:KyotoOkayama}{
     Okayama Observatory, Kyoto University, 3037-5 Honjo, Kamogatacho,
     Asakuchi, Okayama 719-0232, Japan}

\authorcount{affil:OKU}{
     Osaka Kyoiku University, 4-698-1 Asahigaoka, Osaka 582-8582, Japan}

\authorcount{affil:Vanmunster}{
     Center for Backyard Astrophysics Belgium, Walhostraat 1A,
     B-3401 Landen, Belgium}

\authorcount{affil:GEOS}{
     Groupe Europ\'een d'Observations Stellaires (GEOS),
     23 Parc de Levesville, 28300 Bailleau l'Ev\^eque, France}

\authorcount{affil:BAV}{
     Bundesdeutsche Arbeitsgemeinschaft f\"ur Ver\"anderliche Sterne
     (BAV), Munsterdamm 90, 12169 Berlin, Germany}

\authorcount{affil:Hambsch}{
     Vereniging Voor Sterrenkunde (VVS), Oude Bleken 12, 2400 Mol, Belgium}

\authorcount{affil:Ioh}{
     Variable Star Observers League in Japan (VSOLJ),
     1001-105 Nishiterakata, Hachioji, Tokyo 192-0153, Japan}

\authorcount{affil:CrAO}{
     Federal State Budget Scientific Institution ``Crimean Astrophysical
     Observatory of RAS'', Nauchny, 298409, Republic of Crimea}

\authorcount{affil:Monard}{
     Bronberg Observatory, Center for Backyard Astrophysics Pretoria,
     PO Box 11426, Tiegerpoort 0056, South Africa}

\authorcount{affil:Monard2}{
     Kleinkaroo Observatory, Center for Backyard Astrophysics Kleinkaroo,
     Sint Helena 1B, PO Box 281, Calitzdorp 6660, South Africa}

\authorcount{affil:Dvorak}{
     Rolling Hills Observatory, 1643 Nightfall Drive,
     Clermont, Florida 34711, USA}



\maketitle

\begin{abstract}
IW And stars are a recently recognized group of dwarf novae
which are characterized by a repeated sequence of
brightening from a standstill-like phase with damping
oscillations followed by a deep dip.
\citet{kim19iwandmodel} recently proposed a model
based on thermal-viscous disk instability in a tilted
disk to reproduce the IW And-type characteristics.
IM Eri experienced the IW And-type phase in 2018
and we recorded three cycles of the (damping) oscillation phase
terminated by brightening.
We identified two periods during the IW And-type state:
4--5~d small-amplitude (often damping) oscillations and
a 34--43~d long cycle.  This behavior is typical for
an IW And-type star.
The object gradually brightened within the long cycle
before the next brightening which terminated the (damping)
oscillation phase.
This observation agrees with the increasing disk mass
during the long cycle predicted by a model of
thermal-viscous disk instability in
a tilted disk \citep{kim19iwandmodel}.
We, however, did not succeed in detecting negative superhumps,
which are considered to be the signature of
a tilted disk.
\end{abstract}

\section{Introduction}

   Cataclysmic variables (CVs) are binary stars consisting
of a white dwarf and a mass-transferring secondary filling
the Roche lobe.  The transferred matter forms an accretion
disk around the white dwarf.  Some CVs show outbursts
with amplitudes of several magnitudes and they are called
dwarf novae (DNe).  Some CVs do not show prominent outbursts
and stay at almost constant brightness.  They are usually
called novalike variables (NLs).  It has been demonstrated
that thermal instability in the accretion disk
causes outbursts in DNe [see e.g. \citet{osa96review};
for general information of cataclysmic variables
and dwarf novae, see e.g. \citet{war95book}].
In this disk-instability model, objects with sufficiently
high mass-transfer rates ($\dot{M}$) have thermally
stable disks and they are considered to explain NLs without
prominent outbursts.

   Some CVs lie near the border of DNe and NLs.
The best-known class is Z Cam stars, which at times show
DN-type outbursts and also NL-type quasi-constant phases
called standstills.  It is widely believed that
changing $\dot{M}$ from the secondary causes both
DN and NL states in the same object depending on
$\dot{M}$ \citep{mey83zcam}.

   Quite recently, another class of objects (also regarded
as a subclass of Z Cam stars) has been recognized.
The initial two objects, IW And and V513 Cas, showed
similar light curves characterized by brightening from
a standstill (or standstill-like phase) followed by
a deep dip.  These features are known to recur
quasi-periodically \citep{sim11zcamcamp1}.
These two stars were studied in detail by \citet{szk13iwandv513cas}.
\citet{ham14zcam} called these objects
``anomalous Z Cam stars'' and suggested that brightening
from a standstill and a following dip are caused by
an enhancement and a subsequent decay of
$\dot{M}$ from the secondary.  Following \citet{ham14zcam},
these enhancements and decays of $\dot{M}$ need to occur
repeatedly to explain the behavior of these objects.
\citet{ham14zcam} described that ``these [mass-transfer]
outbursts are not found in most systems and may thus be
due to some rare set of circumstances/parameters''
and admitted the difficulty within the current knowledge.

   \citet{kat19iwandtype} identified three more objects
V507 Cyg, IM Eri and FY Vul showing the similar behavior.
Adding two more objects in the literature
[ST Cha \citep{sim14stchabpcra}; KIC 9406652 \citep{gie13j1922}],
\citet{kat19iwandtype} defined the class
having properties of a sequence of a standstill terminated by
brightening followed by damping oscillations.
They are now widely referred to as IW And-type stars
following the suggestion by \citet{kat19iwandtype}.
There was yet another object in the literature
[V523 Lyr \citep{mas16v523lyr}]
which was not included in \citet{kat19iwandtype}.

   The key difference of IW And-type objects from the classical
Z Cam stars can be summarized as follows:
(1) Standstills are terminated by brightening
in IW And-type objects, unlike fading in Z Cam stars.
The termination of standstills in Z Cam stars is usually
understood as a consequence of the decreasing $\dot{M}$
from the secondary [see e.g. \citet{war95book}]
and brightening at the end of a standstill is not expected
from this picture.
(2) In IW And-type objects, there are quasi-periodic cycles
consisting of a (quasi-)standstill with damping oscillations --- 
brightening which terminates the standstill --- often a deep dip
(not always present within the same object) and returning
to a (quasi-)standstill.
Standstills in Z Cam stars occur irregularly and no such cycles
are known.

   \citet{kat19iwandtype} suggested the presence of
a previously unknown type of limit-cycle oscillation
to reproduce the cycle.  They proposed that
a (quasi-)standstill in these objects are somehow
maintained in the inner part of the disk and that
thermal instability starting from the outer part of
the disk sweeps the disk, thereby terminating the standstill
to complete the cycle.  The idea of a limit cycle
is advantageous in the absence of a known mechanism
to produce recurring mass-transfer
outbursts \citep{ham14zcam}.

   In line with this suggestion and using the evidence
of a tilted disk in KIC 9406652 \citep{gie13j1922},
\citet{kim19iwandmodel} proposed an idea that the transferred
matter can reach the inner portion of a tilted disk.
The transferred matter to the inner disk thermally
stabilizes the inner disk, which was expected to
reproduce (quasi-)standstills in IW And-type objects.
Using a one-dimensional thermal-viscous disk instability
model, \citet{kim19iwandmodel} were able to reproduce
a limit-cycle oscillation consisting of standstill-like
damping oscillations terminated by brightening.
This limit cycle reflects the secular increase of
the disk mass during standstill-like damping oscillations,
which is swept by the cooling wave caused by
thermal instability from the outer disk. 
This limit cycle is analogous to the one in SU UMa-type
dwarf novae, in which the secular increase of
the angular momentum plays a similar role \citep{osa96review}.
The existence of a cycle and brightening terminating
the (quasi-)standstill are two key elements
of the IW And-type phenomenon.

\section{IM Eridani}

   Among newly identified IW And-type stars in
\citet{kat19iwandtype}, IM Eri had been known as
an NL variable (\cite{che01ECCV}; \cite{arm13aqmenimeri}).
\citet{arm13aqmenimeri} spectroscopically identified
the orbital period to be 0.1456346(2)~d and found
strong negative superhumps with a period of 0.13841(3)~d
in their 2002 and 2012 observations.
Negative superhumps are semi-periodic variations
having periods slightly shorter than the orbital period
and are considered to arise from a precessing tilted disk
(cf. \cite{woo07negSH}; \cite{mon09negativeSH}).
According to the All Sky Automated Survey (ASAS-3)
observations \citep{ASAS3},
this object was in the NL state between 2000 November
and 2005 April, but apparently showed outbursts between
2005 May and 2009 December.  There were some time-series
observations in the AAVSO database\footnote{
   $<$http://www.aavso.org/data-download$>$.
} between 2011 December and 2016 March, all showing
NL states.  Most recently, the All Sky Automated Survey
for Supernovae (ASAS-SN) Sky Patrol
(\cite{ASASSN}; \cite{koc17ASASSNLC})
started observations since 2013 October, which eventually
led to an identification as an IW And-type star
starting from the dwarf nova-type state since 
2016 September.  Although some of this dwarf nova-type state
was covered by the AAVSO database in 2016--2017,
the data were too sparse to clarify the IW And-type nature.

   After the identification of the ongoing IW And-type state
in IM Eri in 2018 October, we initiated a campaign.
According to \citet{kim19iwandmodel}, even the same object
can display a variety of light variations depending on
the tilt angle of the disk.  The IW And-type state in
IM Eri was expected to provide an excellent opportunity
to study the relationship between the disk tilt and
light variation.

\begin{figure*}
  \begin{center}
    \FigureFile(130mm,90mm){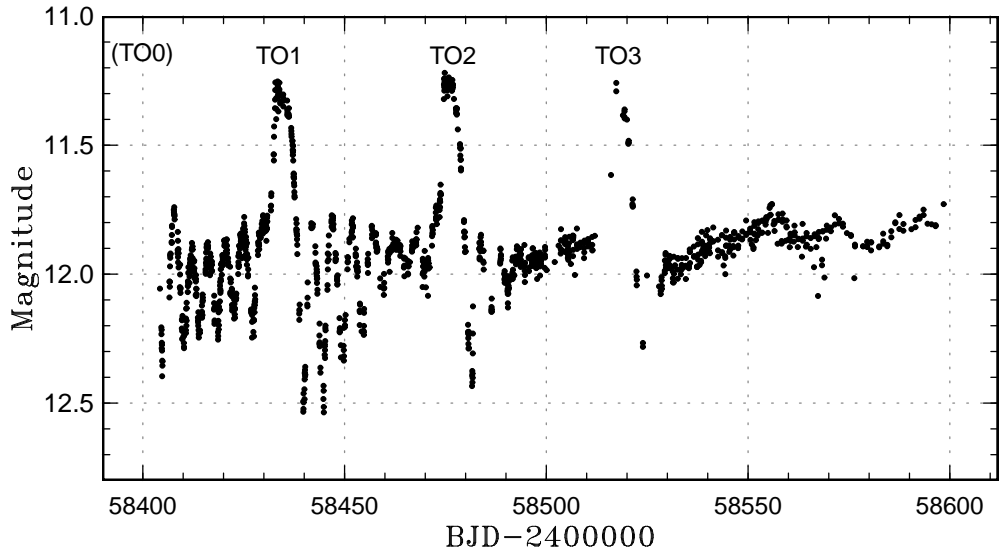}
  \end{center}
  \caption{Light curve of IM Eri in 2018 from our CCD observations.
  The data were averaged to 0.03~d bins.
  We observed three ``terminal outbursts'' (TO1, TO2, TO3)
  at the end of damping oscillations.  The earlier epoch
  of a terminal outburst is shown as ``(TO0)'', which was
  recorded by ASAS-SN, but the data were insufficient to
  depict variations before and after it.
  }
  \label{fig:imerilc}
\end{figure*}

\section{Observation and Analysis}

   The data were acquired by the contributors to
the VSNET Collaboration \citep{VSNET} with 20--60cm
telescopes located world-wide.  All observers used
$V$-filtered or unfiltered CCD cameras.  They used
aperture photometry and extracted magnitudes relative
to comparison stars whose constancy has been confirmed
by comparison with check stars.
The small differences between observers were corrected
by adding constants to minimize the squared
sum of adjacent observations in the combined light curve.
The times of all observations are expressed in 
barycentric Julian days (BJD).
We mainly used R software\footnote{
   The R Foundation for Statistical Computing:\\
   $<$http://cran.r-project.org/$>$.
} for data analysis.
The log of observations and corrections
to individual observers are listed in E-table 1.

   We used the phase dispersion minimization (PDM, \cite{PDM})
method to identify periods in the data.  The error
of the PDM analysis was estimated by the methods
of \citet{fer89error} and \citet{Pdot2}.
Although we also applied the least absolute shrinkage and
selection operator (Lasso) method 
(\cite{lasso}; \cite{kat12perlasso}) particularly using
the two-dimensional version \citep{kat13j1924},
we do show the result since we could not find
particular advantage over the PDM method.

\section{Results}

\subsection{Long-term behavior}

   The resultant light curve is shown in figure \ref{fig:imerilc}.
Our observation started just after the ``terminal outburst'' (TO0),
which was recorded by ASAS-SN.  The object showed oscillations
and gradually brightened.  This oscillation phase was terminated
by the outburst TO1.  We observed three terminal outbursts
TO1, TO2 and TO3 during our campaign.  The intervals between
them were 43~d and 41~d.  The interval between TO0 and TO1
was 34~d, suggesting that the intervals were not perfectly
regular.

   Although the oscillations between TO0 and TO1 were not
particularly damping, they were clearly damping between
TO1 and TO2.  In the segment between TO2 and TO3, the oscillations
were clearly seen with small amplitudes only for
the initial two cycles and gradual brightening toward TO3
remained after them.
The periods of oscillations were 4.34(3)~d in TO0-TO1 and
5.05(3)~d in TO1-TO2.  These periods were also not constant.

\subsection{Orbital and superhump variations}

\begin{figure}
  \begin{center}
    \FigureFile(88mm,70mm){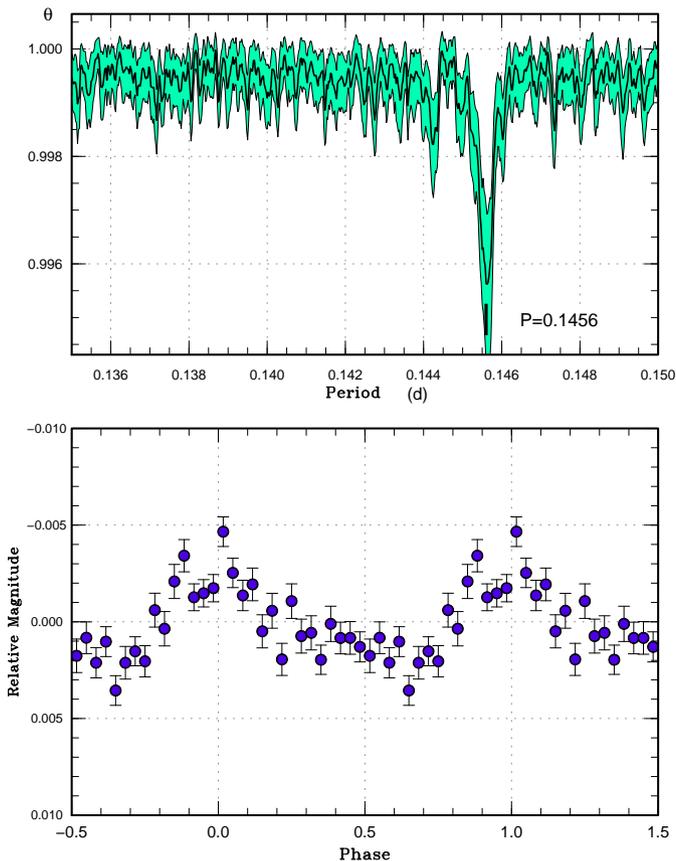}
  \end{center}
  \caption{PDM analysis of IM Eri before BJD 2458474
  (before TO2).
  We analyzed 100 samples which randomly contain 50\% of
  observations, and performed the PDM analysis for these samples.
  The bootstrap result is shown as a form of 90\% confidence intervals
  in the resultant PDM $\theta$ statistics.}
  \label{fig:imeripdm12}
\end{figure}

   We show the results of the PDM analysis of (damping)
oscillation phases (including TO1) before TO2 in
figure \ref{fig:imeripdm12}.
To remove the trend of oscillations, the original data
were cut to 1-d segments and de-trended using
the locally-weighted polynomial regression (LOWESS: \cite{LOWESS})
when the data were abundant, and using a linear regression
when the data number of each bin was less than 50.
The result was virtually the same when we treated
longer segments by LOWESS only.  The result was also
virtually the same when we included/excluded
terminal outbursts in the segment of analysis.
No signal was detected between TO2 and TO3, probably
due to the smaller number of observations.

   The data showed only a weak signal of the orbital
period at 0.1456(1)~d, which is identical with
the spectroscopic period of 0.1456346(2)~d in
\citet{arm13aqmenimeri}.  There was no signal
of negative superhumps near the expected region
around 0.1384~d.  To check the effect of de-trending,
we also performed the PDM analysis without de-trending
or weaker de-trending by changing LOWESS parameters.
No signal of negative superhumps appeared regardless
of the degree of de-trending.  We also applied analysis
for shorter segments (E-figures 1,2) without detectable 
signal of negative superhumps.
We conclude that the amplitude
of negative superhumps was below the detection limit
(0.01 mag).

\begin{figure}
  \begin{center}
    \FigureFile(88mm,70mm){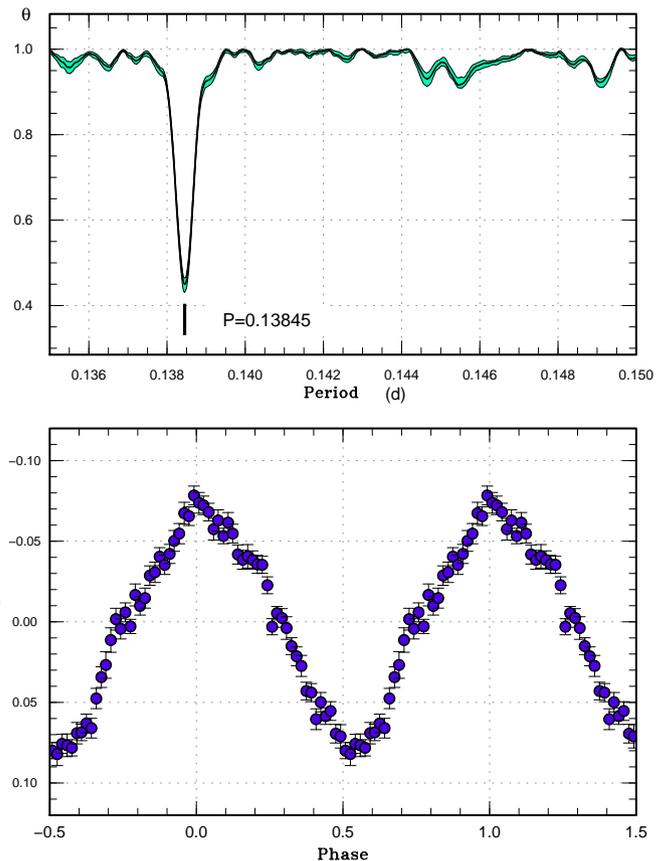}
  \end{center}
  \caption{PDM analysis of IM Eri in the NL state
  in 2012.  Negative superhumps were very strong.
  The orbital signal at 0.1456~d was much weaker than
  the signal of negative superhumps.}
  \label{fig:imeri2012}
\end{figure}

   The result was very different from the PDM analysis
of the 2012 data [a major part of the data used
in \citet{arm13aqmenimeri} is available in
the AAVSO database] when the object was
in the NL state and negative superhumps were very
strong (amplitude of 0.16 mag, figure \ref{fig:imeri2012}).

\section{Discussion}

\subsection{Overall behavior}

   We found that IM Eri showed two periods
during the IW And-type state: 4--5~d small-amplitude
oscillations (often damping) and a 34--43~d long
cycle.  This behavior is typical for an IW And-type star.

   We first examine whether these small-amplitude
oscillations reflected the intrinsic variation of
the disk luminosity or whether they were caused by
the varying aspect of a precessing tilted disk.
Even though there remains a possibility of a tilted
disk producing undetectably weak negative superhumps,
the beat period between the orbital period and the period
of recorded negative superhumps in IM Eri is 2.8~d,
which cannot explain the 4--5~d small-amplitude
oscillations.  The small-amplitude
oscillations thus should arise from the intrinsic
variation of the disk luminosity.

   In the case of the Kepler object KIC 9406652,
the long-term variation was strongly affected by
the beat period between the orbital period and
the period of negative superhumps
\citep{gie13j1922} and it is difficult to unambiguously
distinguish damping oscillations and the beat variation
(the full treatment of this issue considering the variable
precession rate will be discussed in Kimura et al. in prep.).
Since the orbital periods and periods of negative
superhumps, if present, were unknown or only poorly known
in other IW And-type stars, the distinction between
damping oscillations and the beat phenomenon is even
more difficult.  IM Eri thus provides the first clear
demonstration of the (damping) oscillations between
terminal outbursts by time-resolved photometry.

   There was a gradual brightening tendency during
the (damping) oscillation phase.  This gradual brightening
cannot be explained straightforwardly by recurrent
mass-transfer bursts as proposed by \citet{ham14zcam}.
This tendency, however, can be well explained by the gradual
increase of the disk mass by the model of a tilted disk
by \citet{kim19iwandmodel}.

   After TO3, this object gradually brightened without
oscillations.  (Terminal) outbursts also disappeared.
It was likely that $\dot{M}$ gradually increased after TO2:
the outer part of the disk gradually became
thermally stable, which suppressed thermal instability
producing a terminal outburst to occur.

\subsection{Evidence for disk tilt}

   In contrast to the 2012 observations,
IM Eri did not show a strong sign of negative superhumps,
which is the signature of a disk tilt, during the IW And-type
state.  This is against what would be expected by
\citet{kim19iwandmodel}.  The 2012 NL state
was probably indeed achieved by a sufficient increase
of $\dot{M}$, and the existence of the disk tilt
did not affect the condition of thermal stability
throughout the disk.  It might be possible that a weak tilt
(with a possible signal of negative superhumps below
the detection limit) was present in 2018
when $\dot{M}$ somewhat decreased.

   We must note that our observations were not as ideal
as in KIC 9406652 and had long longitudinal observational gaps.
We had only 6~hr (at most) continuous runs each night
with samplings in every 4--5~min, and this observing
condition was not ideal for detecting a weak signal
of negative superhumps.  A search for negative superhumps
with a better temporal coverage and resolution when
IM Eri is again in the IW And-type state, or in other objects,
would solve the current discrepant result.

\section*{Acknowledgments}

We acknowledge with thanks the variable star
observations from the AAVSO International Database contributed by
observers worldwide and used in this research.
We are also grateful to the ASAS-3 and ASAS-SN teams for making
the past photometric database available to the public.

\section{Summary}

   We made time-resolved photometry of IM Eri, which experienced
the IW And-type state (a repeated sequence of standstill-like
damping oscillation terminated by brightening followed
by a dip) in 2018.  We recorded three cycles and
following the brightening in 2019 February.
After these three cycles, the object gradually brightened to
a standstill which lasted for more than 70~d.
We identified two periods during the IW And-type state:
4--5~d small-amplitude oscillations (often damping) and 34--43~d long
cycle.  We confirmed that these small-amplitude oscillations
cannot be explained by the changing aspect of a precessing
tilted disk.  The states of small-amplitude oscillations
were terminated by ``terminal outbursts''.
The entire behavior is typical for an IW And-type object
and the current observation provides the best confirmation
of damping oscillations arising from the intrinsic variation
of the disk luminosity.  The object gradually brightened between these
terminal outbursts.  This finding agrees with the prediction
of thermal-viscous disk instability in
a tilted disk \citet{kim19iwandmodel}.
We, however, did not succeed in detecting negative superhumps,
which are considered to be the signature of a tilted disk.

\section*{Supporting information}
Additional supporting information can be found in the online version
of this article: E-table 1, E-figures 1,2.
Supplementary data is available at PASJ Journal online.

\end{document}